# Role of $E_V$+0.98 eV trap in light soaking-induced short circuit current instability in CIGS solar cells

P. K. Paul[1], T. Jarmar[2], L. Stolt[2], A. Rockett[3], and A. R. Arehart[1]

[1]*Electrical and Computer Engineering, The Ohio State University, Columbus, OH USA*
[2] *Solibro Research AB, Uppsala, Sweden*
[3]*Department of Metallurgical and Materials Engineering, Colorado School of Mines, Golden, CO USA*

*Abstract* — Light-induced instabilities/degradation in Cu(In,Ga)Se$_2$ (CIGS) solar cells are a prevalent and urgent issue to resolve to improve performance, uniformity, and reliability. Here, mechanisms contributing to light–induced instabilities are identified focusing on an observed short circuit current (J$_{SC}$) reduction. External quantum efficiency measurements before and after light soaking identified a reduction in long wavelength photon carrier collection efficiency in the CIGS absorber layer. Using deep level optical spectroscopy (DLOS), the concentration of CIGS E$_V$+0.98 eV deep level is correlated with the amount of J$_{SC}$ degradation, Finally, capacitance voltage (C-V) measurements reveal light induces a large reduction in the depletion depth and reduction of carrier collection and are all correlated with the J$_{SC}$ reduction. Finally, the E$_V$+0.53 eV trap concentrations are shown to correlate with V$_{OC}$ instability but not the J$_{SC}$ reduction confirming that multiple trap-induced mechanism are responsible for the light-induced instabilities.

## I. INTRODUCTION

CIGS is an established absorber material for thin-film solar cells due to its high optical absorption, tunable bandgap, and low manufacturing cost [1]. The record efficiency of CIGS thin film solar cell is 22.6%, which is the highest among all thin-film solar [2]. In spite of high initial efficiency, CIGS solar cell instabilities due to the effects of light, temperature and moisture is an active research area [3]. Previously, CIGS solar cells have shown both beneficial and detrimental effects after extended light soaking (LS) [4]. Typically, CIGS solar cells subjected to LS exhibit increased effective p-type doping due to changes in trap occupancy [5]. Additionally, deep levels and interface traps are believed to be responsible for loss of solar cell performance [6], but the actual mechanisms and the knowledge about the individual traps responsible for LS-induced degradation are still not well understood. In CIGS, light-soaking can impact open circuit voltage (V$_{OC}$), Jsc fill factor (FF), and efficiency (η), but since it influences so many cell parameters it suggests more than one mechanism is responsible for all these changes [7]. To understand the mechanisms of LS- induced instability, we investigated solar cells with different LS responses.

## II. APPROACH

In this study, three CIGS solar cells with varying degrees of instability grown by Solibro Research AB were used. First, Mo metal back contacts were deposited on a soda lime glass substrate and then the CIGS absorber layer was deposited by a 2-stage vacuum co-evaporation process. Then the CdS buffer layer was deposited by chemical bath deposition. The front contact was formed by sputtering intrinsic ZnO (IZO) and Al-doped ZnO (AZO) layers. The cells were characterized then ~1 mm$^2$ sub-cells were physically circumscribed to provide devices with suitable capacitance for the capacitance-based measurements.

To understand and identify the role of deep traps responsible for the light-induced J$_{SC}$ degradation, capacitance–voltage (C-V) profiling and deep level transient and optical spectroscopies (DLTS/DLOS) were all performed with external quantum efficiency (EQE) measurements. The DLTS transients were analyzed using the double boxcar method with rate windows from 0.8 to 2000 s$^{-1}$, and the equipment and additional details are described in Ref. 8. For the DLTS and DLOS measurements, the traps were filled with 10 ms and 10 s with +0.4 V pulses, respectively, and the trap emission was recorded at -1.0 V. The DLOS capacitance transients were recorded for 350 s at room temperature with monochromatic light incident with photon energies from 0.5 to 1.4 eV in 0.02 eV steps. C-V measurements were performed using a Boonton 7200 capacitance meter and the net doping profiles (*N*) were calculated using [9]

$$N = \frac{-C^3}{q\varepsilon_s A^2 \left(\frac{dC}{dV}\right)}$$

where *A* is the device area, ε$_s$ is the permittivity, and *q* is the elementary charge.

## III. RESULTS AND DISCUSSION

Table I shows the impact of light soaking at 1000 W/m$^2$ at approximately 25$^0$ C for 24 h on the three samples. After light soaking, Sample 1 shows the smallest J$_{SC}$ reduction and Samples 3 show ~6X larger J$_{SC}$ reduction. Additionally, Samples 1 and 2 show reduced open circuit voltage (V$_{OC}$), which is a different pattern of instability than the J$_{SC}$ reduction. To understand the light-induced J$_{SC}$ reduction, EQE measurements were performed before and after LS. From Fig. 1, the EQE measurements show a clear decrease in carrier collection efficiency at longer wavelengths after LS. The longer wavelength photons are absorbed deeper in the sample (i.e. in the CIGS absorber layer), this indicates that the CIGS layer is responsible for the loss of carrier collection. The carrier collection efficiency loss is clearly correlated with the measured LS-induced $J_{SC}$ reduction. The collection efficiency is dictated by the minority carrier diffusion length $L_n$ and the depletion depth $W$, which is given by [9]

$$W = \sqrt{\frac{2\varepsilon_s(V_{bi}-V)}{qN_A}}$$

where $V$ is the applied voltage, $V_{bi}$ is the built-in potential, and $N_A$ is the carrier concentration. In the case of CIGS, the trap densities can be larger than the doping concentration, which results in large changes in the depletion depth, and hence is a likely source of the reduced carrier collection efficiency.

TABLE I
Light-induced changes for the three CIGS samples

|  | Sample 1 | Sample 2 | Sample 3 |
|---|---|---|---|
| η change (%) | -0.6 | -1.16 | -0.6 |
| V$_{OC}$ change (V) | -0.01 | -0.02 | +0.005 |
| **J$_{SC}$ change (mA/cm$^2$)** | **-0.20** | **-0.85** | **-1.25** |
| FF change (%) | -1.4 | -1.8 | 0.0 |
| [E$_V$ + 0.98 eV] (cm$^{-3}$) | 1.5×10$^{15}$ | 5.0×10$^{15}$ | 6.0×10$^{15}$ |
| C-V N$_A$ change (cm$^{-3}$) | 1.0×10$^{15}$ | 3.5×10$^{15}$ | 5.1×10$^{15}$ |
| [E$_V$ + 0.53 eV] (cm$^{-3}$) | 1.3×10$^{13}$ | 2.3×10$^{13}$ | BD |
| BD = below detection | | | |

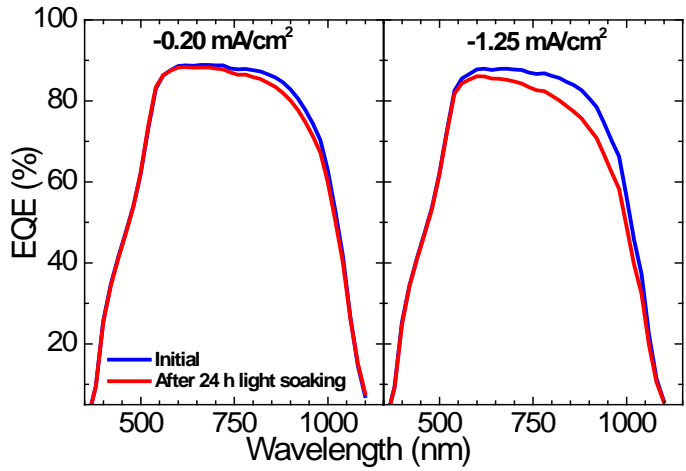

**Figure 1**: EQE spectra before and after light soaking (left) Sample 1 (-0.20 mA/cm$^2$) with the lowest instability and (right) Sample 3 (-1.25 mA/cm$^2$ $J_{SC}$ degradation) with the highest instability. Larger reduction in EQE is observed in the long wavelengths of Sample 3, consistent with the $J_{SC}$ degradation and indicating the CIGS layer is responsible for the $J_{SC}$ reduction.

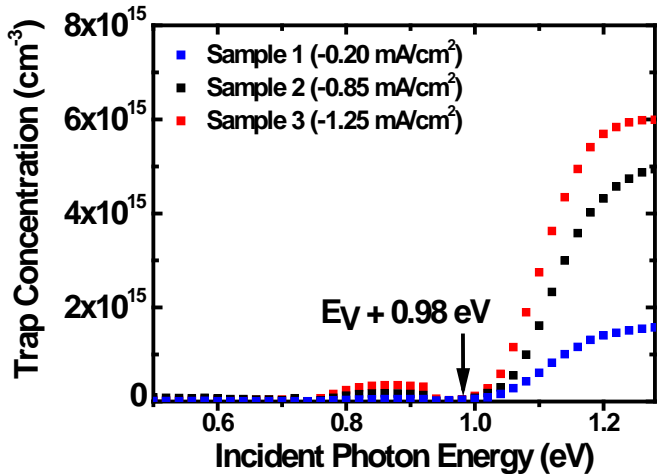

**Figure 3:** DLOS steady state photocapacitance spectra showing a correlation between the $E_V$+0.98 eV concentration and $J_{SC}$ reduction. The trap concentrations are summarized in Table I.

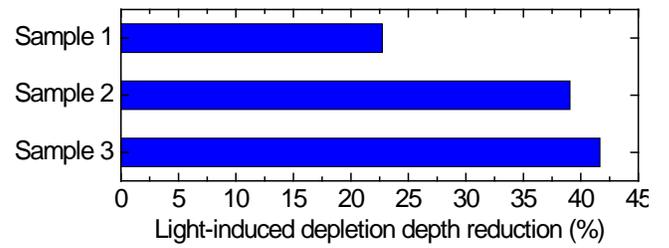

**Figure 2:** Depletion depth reduction for Sample 1 (-0.2 mA/cm$^2$), Sample 2 (-0.85 mA/cm$^2$), and Sample 3 (-1.25 mA/cm$^2$) after–LS soaking. The samples with higher $J_{SC}$ instability have the highest relative reduction of the depletion region leading to higher $J_{SC}$ reductions.

To understand the $J_{SC}$ reduction likely due to the trap-induced depletion depth changes, C-V measurements were performed: in the dark to maintain the equilibrium trap occupancy (e.g. traps filled with holes), and with 1.3 eV illumination light to empty all traps in the bandgap. Because the monochromatic light intensity (~10$^{14}$-10$^{15}$ photons/cm$^2$/s) is much less than the typical above bandgap solar spectra, any changes observed under this illumination will also be observed under normal solar cell operation. From the C-V measurements, the carrier concentration vs. depth was extracted [9]. From Fig. 2, LS caused a significant reduction in the depletion width for all three samples. Importantly, after LS Samples 2 and 3, which exhibit the highest $J_{SC}$ reduction, have the highest relative change of the depletion width where after LS Samples 2 and 3 have a depletion depth (0.35-0.39 μm) that is less than half that of Sample 1 (0.85 μm). Thus, the light-induced depletion depth reduction is the likely source of the carrier collection efficiency reduction, reduction in the long wavelength EQE, and also the reduction in $J_{SC}$. The precise source of this depletion depth reduction can be easily identified in the defect spectroscopy measurements.

The DLOS measurements in Fig. 3 show a trap at $E_V$ + 0.98eV in all samples, which was previously suggested to be the $V_{Cu}+V_{Se}$ defect [10]. Table I shows a clear trend between the increase in effective doping measured by C-V (Fig. 2) due to 1.3 eV illumination and the concentration of the $E_V$+0.98 eV level indicating the $E_V$+0.98 eV trap is likely responsible for the light-induced $J_{sc}$ reduction in these samples and that reduction of the $V_{Cu}+V_{Se}$ divacancy concentration would be necessary to alleviate the loss of $J_{SC}$.

To complete the defect study throughout the CIGS bandgap, DLTS measurements were performed. Fig. 4 shows DLTS measurement spectra where a trap at $E_V$+0.53 eV is present in all samples. The trap concentrations are listed in Table I, but it is evident that there is no clear correlation between the trap concentration and light-induced $J_{SC}$ reduction. However, there is a clear correlation between the $V_{OC}$ change and the $E_V$+0.53 eV concentration. Because the $E_V$+0.53 eV trap is near mid-gap, it might be an efficient recombination center and efficiently reduce $V_{OC}$, but this mechanism is still under investigation.

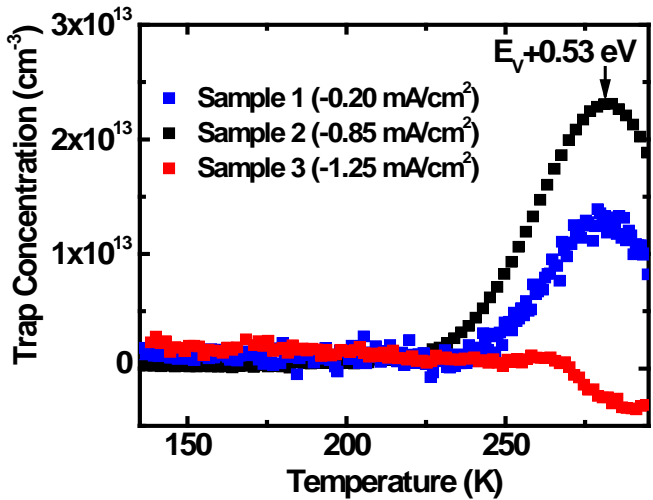

**Figure 4**: DLTS spectra showing trap with activation energy $E_V+0.53$ eV. Concentration of $E_V+0.53$ eV shows no correlation with $J_{SC}$ reduction indicates $E_V+0.53$ eV might not be the cause of $J_{SC}$ reduction.

### IV. CONCLUSIONS

CIGS samples subjected to light soaking results in changes in solar cell performance that are correlated with the $E_V+0.98$ eV level in the case of the observed $J_{SC}$ reduction and $E_V+0.53$ eV level for the $V_{OC}$ changes. The $J_{SC}$ reduction results from loss of carrier collection in the CIGS layer where the $E_V+0.98$ eV level causes a large reduction in the depletion depth when light above 1.0 eV is incident on the sample (i.e. normal solar cell conditions), which is the likely source of the $J_{SC}$ reduction. On the other hand, the $V_{OC}$ change is correlated with the concentration of the near mid-gap $E_V+0.53$ eV level, which may be an efficient recombination center. This suggests there is indeed more than one mechanism for the light-soaking-induced changes in cell performance and reducing the concentration of these two deep levels would help mitigate the light-soaking induced $J_{SC}$ and $V_{OC}$ instabilities.

**Acknowledgements**

The authors would like to thank the Department of Energy (Contract #DE-DD0007141) for partial financial support.